\documentclass{cimento}
\usepackage{graphicx}        % For eps figures, newer & more powerfull
\usepackage{color}           % For color text: \color command
\usepackage{url}             % For breaking URLs easily trough lines
\title{A new approach for the heliometric optics}

\author{V. d'Avila,\from{ins:e} E. Reis Neto,\from{ins:m} A. Coletti,\from{ins:c} L. C. Oliveira,\from{ins:o} V. Matias,\from{ins:o} A. H. Andrei,\from{ins:e} J. L. Penna,\from{ins:e} S. Calderari Boscardin,\from{ins:e} C. Sigismondi, \from{ins:i}\from{ins:e}}

\instlist{
\inst{ins:e} Observat\'orio Nacional/MCTI, Rio  de Janeiro (Brasil)
\inst{ins:m} Museu de Astronomia/MAST/MCTI, Rio  de Janeiro (Brasil) 
\inst{ins:c} Azeheb/FACC, Curitiba (Brasil) 
\inst{ins:o} NGC-51, Rio  de Janeiro (Brasil)
\inst{ins:i} ICRA, International Center for Relativistic Astrophysics, Sapienza University of Rome, P.le Aldo Moro 5 00185, Roma (Italy); Universit\'e de Nice-Sophia Antipolis (France); IRSOL Istituto Ricerche Solari di Locarno (Switzerland)}

\PACSes{\PACSit{42.15.Eq}{Optical system design}
\PACSit{95.55.Ev}{Solar instruments}
\PACSit{96.60.-j}{Solar physics}
\PACSit{96.60.Bn}{Solar Diameter}}

\begin{document}

\maketitle

\begin{abstract}

The heliometer of Fraunhofer in Koenigsberg (1824) is a refractor in which the lens is split into two halves to 
which is applied a linear displacement along the cut. Later in 1890s a variation of the heliometer 
has been realized in Goettingen using a beam splitting wedge: these methods were both subjected to chromatic and refractive aberrations;
the second configuration being much less affected by thermal fluctuations.
The reflector version of the heliometer conceived at the Observatorio Nacional of Rio de Janeiro
overcome these problems: the two halves of the vitrified ceramic mirror split at a fixed heliometric angle
produce the two images of the Sun exempt of chromatisms and distortions.
The heliometer of Rio is a telescope which can rotate around its axis, to measure the solar diameter
at all heliolatitudes.
A further development of that heliometer, now under construction, is the annular heliometer, 
in which the mirrors are concentric, with symmetrical Point Spread Functions.
Moreover the location of the Observatory of Rio de Janeiro allows zenithal observations, 
with no atmospheric refraction at all heliolatitudes, in December and January.

\end{abstract}

\section{Introduction}

Recent research on global climate changes points to three distinct 
sources of climate disturbance: anthropogenic; natural changes in the 
oceans and atmosphere; and irregularities in the solar cycles. One of 
the most direct way to survey the last origin of climatic variability 
is through the measurement of variations in the diameter and the shape 
of the solar disk. The heliometric method is one of the most successful 
techniques to measure small variations of angles. Its principle has been
used for the latest space borne astrometric missions, aiming to milli-arcsecond precision.\cite{Sigismondi11} The success of this method is in the fact 
that it minimizes the dependence of angular measurements to the thermal 
and mechanical stability of the instrument. It follows that the classic 
heliometer is a refractor in which the lens is split into two halves to 
which is applied a linear displacement along the cut (see Fig. 1). However, this 
classical configuration leaves still room for a residual dependence with 
the focus, arising from due to non-concentricity of the beams of the two 
images. An improvement of the Fraunhofer Heliometer (1824) was obtained 70 years later 
with the Goettingen heliometer,\cite{Ambronn} reproduced in the optical configuration of 
the balloon borne telescope SDS.\cite{Chiu} Since solar observations are by nature subject to large 
temperature variations, the problem was tackled on its basis in the
making of the Solar Heliometer of Observatorio Nacional.\cite{IAUS09} 
The mirrors and its niche are made in CCZ, a vitrified ceramic, 
and the tube on carbon fiber, both materials of negligible thermic coefficients.

Temperature and barometric pressure are determined in real time and 
associated to each measure.

\begin{figure}
\centerline{\includegraphics[width=1\textwidth,clip=]{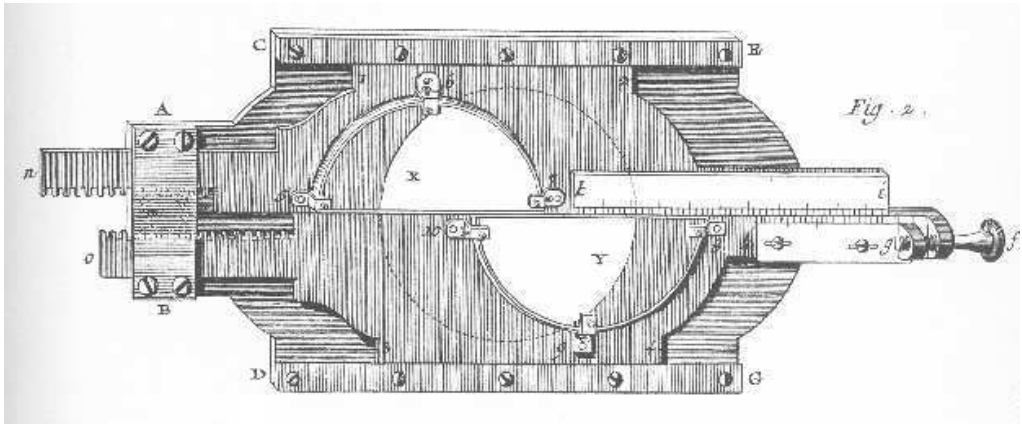}}
\caption{The {\it English Heliometer} in the Encyclopedia of Diderot et D'Alembert.\cite{Diderot}}
\label{Fig. 1}
\end{figure}

\section{Monitoring solar variability}

Stars are kept in place as a product of the balance between the outward
radiation pressure produced by atomic fusion in the core and the inward
force produced by the sheer weight of the unbound layers of plasma. The
surface drawn by such hydrostatic equilibrium defines the stellar boundaries.
Thus the volume (or radius) in itself or combined with additional information
gives way important information concerning the mass, temperature, age and
metallicity of the star. Departures of the surface of hydrostatic equilibrium from a spherical
volume tell of stellar magnetic field, activity cycles and sub-surface
pulsation modes. In particular the measurement of
the solar radius is important when combined to other solar activity,
space weather, and long term Earth climate parameters.

Measurements of the Solar diameter have been made from historical times.
Among the earlier series the more complete and consistent are those issued
from Mercury transits\cite{Sveshnikov} and eclipses,\cite{Sigismondi11} and thes from the observatories of Greenwich and Campidoglio (Rome).\cite{Gething} 

From the second half of the last century, the number of measures increased. Yet, the results are still unsure. 
Variations of the diameter ranging 1 arcsec were
reported, but seldom confirmed. Smaller variations at the level of tens
of milliarcsecond (up to few hundreds of milliarcsecond over the 11 years
cycle) were derived by solar astrolabes, and are compatible also with the
extrapolation of solar measurements from space. Nevertheless when
comparing with the solar activity, both an in-phase and an out-of-phase
signatures befit the data. The semi-diameter is seen to vary in-phase with
solar activity for short periods and high bursts, but off- or out-of-phase
when long periods regarding the 11 year cycle are considered.

At Brazil's Observatorio Nacional (ON) in Rio de Janeiro
regular measurements of variations of the solar diameter started in 1997
with a CCD astrolabe.\cite{Neto2} The series extended till 2009 with a very high density
of daily measures (from 10 up to 30). In 2010-11 the instrument is being upgraded and the service is interrupted.\cite{Boscardin}

From 2008 onwards, and since 2009 with an important
subvention of Brazil's science foundation FINEP, an Heliometer started
to be developed and build at ON (see Fig. 3).\cite{Neto} The instrument had its first light in 2009, and entered in routine operation by the end of 2010 after 
a field test, during the 2010 total solar eclipse at Easter Island.

\begin{figure}
\centerline{\includegraphics[width=1\textwidth,clip=]{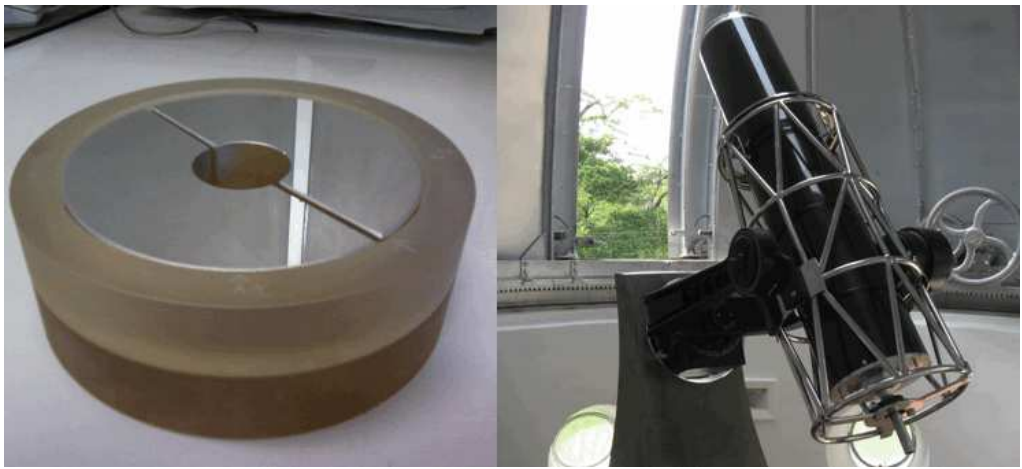}}
\caption{The Heliometer now operating in Rio de Janeiro, with its splitted mirror-objective.}
\label{Fig. 2}
\end{figure}

\section{Solar astrolabes and heliometer's concepts in comparison}

Solar astrolabes have provided continuous results of the variations of the
solar diameter since 1975. It started with the observations of F. Laclare at
the Calern Station of the Observatoire de la Cote d'Azur. Immediately followed
similar observations at the Instituto Astronomico e Geofisico of the
Universidade de Sao Paulo, at the Observatorio de Chile, and at the Real
Observatorio de la Armada in San Fernando, Spain. In the later years of the
XXth century the Observatorio Nacional also started its observations,
immediately followed by the Antalya station of the Akdeniz University, Turkey.
Finally, a few years later an improved instrument, based on the astrolabe
principles, the DORaySol started operations at the same Calern Observatory.\cite{Laclare}

The results from all those settlements show a remarkable coherence, eventhough
as previously mentioned several open questions remained. Fig. 3 presents
the compound series of solar astrolabes highlighting the observed variation of
the diameter.

\begin{figure}
\centerline{\includegraphics[width=1\textwidth,clip=]{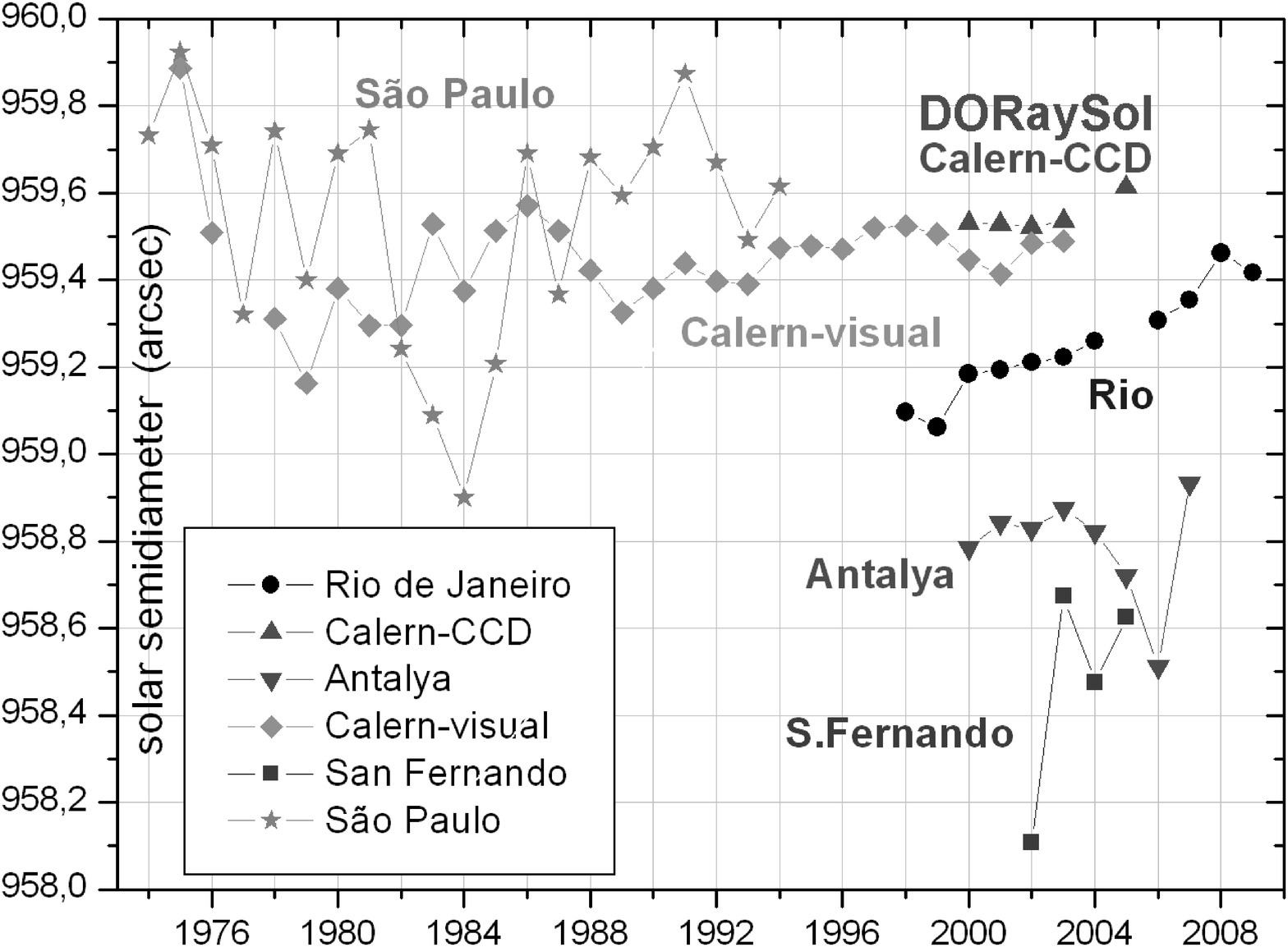}}
\caption{Solar semidiameter yearly averages by R2S3 (R\'eseau de Suivi du Rayon Solaire au Sol) astrolabes series.\cite{Boscardin}}
\label{Fig. 3}
\end{figure}

The lengthy success of the solar astrolabes is based on the remarkable
metrological qualities of the original Danjon astrolabe. The
Danjon astrolabe was originally developed for the field campaigns of
International Geophysical Year of 1958.\cite{Danjon} It was originally used fort stellar observations, defining an
almucantarat and hence the local zenith. Variations of the observed zenith could
be translated into variations of the polar axis direction and spinning of the Earth. 

Three
aspects make the measurements relative, getting rid of instrumental and
observational biases. 

\begin{itemize}
\item{1. The transit through the almucantarat was defined by the
instant of contact of the mirrored images of the star;}

\item{ 2. the time of contact
itself was determined from a collection of measures symmetric in height and
azimuth; and}

\item{ 3. the almucantarat itself was defined by a collection of stars of known catalog position. For the solar observations such conditions were
naturally improved.}
\end{itemize}

The observation becomes twice relative since the upper and
lower solar limbs are measured in sequence; the number of symmetric points
relatively to the almucantarat could be increased at will, and placed closer to
the point of contact of the mirrored images since the acquisition became
digital; the exact placement and homogeneity of the almucantarat itself did not
matter anymore; the several measurements became fully independent, each one
representing a complete determination of the solar diameter; the ancillary
quantities of exact coordinates of the station and clock offset did not matter
anymore either.

The concept of the heliometer is entirely opposite, and so it is its
instrumental assembling. Conceptually it is an absolute measure, in which
an agle is confronted against an instrumental standard. Moreover, the
angle to be measured is small (the variation of the solar diameter)
confronted with the corresponding linear displacement at the focal plane,
thus an error on the linear measurement is smaller by orders of
magnitude over the angular variation that is being measured. Additionally,
if the plate scale can't be determined, the distance between two given
points depends on where lays the image, that is there is a focus dependence.
In practice for the solar observations the effect is much further minimized
because the plate scale is instantaneously know by timing the solar
movement over the detector.

At the Observatorio Nacional, after testing several options,\cite{Neto2} the heliometer
was build through bisecting mirror. Each half-mirror is tilted of an angle slightly greater than 0.135 degrees in order
to displace the images relatively to each other by one solar diameter approximately. In
this way we will have opposite limbs of the Sun almost in tangency in the focal plane at the  perihelion.
The heliometric mirror is all made of CCZ-HS, a ceramic material with very low thermal 
expansion coeffcient ($0.0 \pm 0.2 \times 10^{-7}/^{o}C$). The two half mirrors are immobilized,
in relation to each other, by means of a external ring, all resting over an optical plate.
Its cell guarantees the mechanical and geometrical stability for the entire set.
The mask seen at the top of the cell has been designed to keep the two half-
mirrors blocked in place and also assures that entrance pupil has a symmetric shape.
The surface quality of the optical plate and the mirrors is better than 1/12 and $1/20 \lambda$,
respectively.

As already said, the tube of the telescope is made of carbon fiber. This material is extremely rigid and has very 
low coefficient of thermal expansion. It is mounted inside a stainless steel
truss support and can rotate around its axis.
In order to eliminate the secondary mirror the CCD chip was removed apart from the
camera electronics and installed directly in the focal plane, on a support also made of carbon fiber.
In table 1 the number of observations made with the astrolabe is compared with the observations made with the heliometer.
In 6 months with the heliometer the number of observations made has been 10 times larger than using the heliometer in 12 years, with a rate 240 times larger, and a lower statistical dispersion.
\begin{table}
  \caption{Heliometer vs Astrolabe in Rio de Janeiro National Observatory.}
  \label{tab:Plan}
  \begin{tabular}{rcl}
    \hline
      Period    & Number of observations  & standard deviation [arcsec]   \\
      Astrolabe 1998-2009 & 21640 & 0.590    \\
      Heliometer July 2011 & 54731 & 0.238    \\
      August 2011 & 21767 & 0.213    \\
      September 2011 & 31343 & 0.195    \\
      October 2011 & 54747 & 0.223    \\
      November 2011 & 32661 & 0.274    \\
      December 2011 & 44797 & 0.180    \\
      Total Heliometer 6 months & 240046 &     \\
    \hline
  \end{tabular}
\end{table}

\section{Forthcoming improvements: the annular heliometer}

Simultaneously to the numerous solar observations made nowadays with the Heliometer, an optical system improvement is underway: the new mirror-objective was cut in form of concentric rings. Between them an angular displacement is applied in order to generate two solar images, as the current instrument mirror-objective. With this new instrument, the second generation of the reflector heliometer under construction, it will be possible to achieve the ideal conditions of the heliometric technique, in the sense that, for the first time, angular measurements become totally independent on the instrumental focus stability. By this unique and simple optical characteristic, this telescope is suitable for space missions.

%%% BIBLIOGRAPHY %%%%%%%%%%%%%%%%%%%%%%%%%%%%%%%%%%%%%%%%%%%%%%%%%%%%%%%%%%%
%\mbox{}~\\ 

% name your Bibtex file containing your references (.bib)

\bibliography{helio}
   
\bibliographystyle{varenna}

\end{document}